\documentclass[journal]{IEEEtran}

\ifCLASSINFOpdf
\else
\fi

\usepackage{amsmath,amsfonts}
\usepackage{amsthm}
\usepackage{algorithmic}
\usepackage{verbatim}
\usepackage{graphicx}
\usepackage{cite}
\usepackage[ruled,linesnumbered]{algorithm2e}
\usepackage{cancel}
\usepackage{indentfirst}
\usepackage{epstopdf}
\usepackage{caption}
\usepackage{color}
\usepackage{subfigure}
\usepackage[colorlinks,citecolor=blue,linkcolor=blue,urlcolor=blue]{hyperref}
\usepackage{cite}

\begin{document}
	
	\title{The Future is Fluid: Revolutionizing DOA Estimation with Sparse Fluid Antennas}
	\author{He Xu, Tuo Wu, Ye Tian, Ming Jin, Wei Liu, Qinghua Guo, Maged Elkashlan, \\Matthew C. Valenti, \emph{Fellow}, \emph{IEEE}, 	Chan-Byoung Chae, \emph{Fellow, IEEE},  \\Kin-Fai Tong,  \emph{Fellow, IEEE}, and  Kai-Kit Wong, \emph{Fellow}, \emph{IEEE}  
		\thanks{ (\textit{Corresponding author: Tuo Wu.})
			
			H. Xu is with the School of Cyber Science and Engineering, Ningbo University of Technology, Ningbo 315211, China (E-mail: $\rm xuhebest@sina.com$).
			T. Wu is with the School of Electrical and Electronic Engineering, Nanyang Technological University 639798, Singapore (E-mail: $\rm  tuo.wu@qmul.ac.uk$).
			W. Liu is with the Department of Electrical and Electronic Engineering, Hong Kong Polytechnic University, Kowloon, Hong Kong (E-mail: $\rm wei2.liu@polyu.edu.hk$).
			Y. Tian and M. Jin are with the Faculty of Electrical Engineering and Computer Science, Ningbo University, Ningbo 315211, China (E-mail: $\rm tianye1@nbu.edu.cn, jinming@nbu.edu.cn$).
			Q. Guo is with the School of Electrical, Computer, and Telecommunications Engineering, University of Wollongong, Wollongong, NSW 2522,
			Australia (E-mail: $\rm qguo@uow.edu.au$).
			M. Elkashlan is with the School of Electronic Engineering and Computer Science at Queen Mary University of London, London E1 4NS, U.K. (E-mail: $\rm maged.elkashlan@qmul.ac.uk$). 
			M. C. Valenti is with the Lane Department of Computer Science and Electrical Engineering, West Virginia University, Morgantown, USA (E-mail: $\rm valenti@ieee.org$).
			K. F. Tong is with the School of Science and Technology, Hong Kong Metropolitan University, Hong Kong SAR, China. (E-mail: $\rm ktong@hkmu.edu.hk$).
			C.-B. Chae is with the School of Integrated Technology, Yonsei University, Seoul 03722 Korea. (E-mail: $\rm cbchae@yonsei.ac.kr$).
			K.-K. Wong is with the Department of Electronic and Electrical Engineering, University College London, WC1E 6BT London, U.K., and also with the Yonsei Frontier Laboratory, Yonsei University, Seoul 03722, South Korea (E-mail: $\rm kai$-$\rm kit.wong@ucl.ac.uk$).
	}}
	
	\markboth{IEEE Transactions on Wireless  Communications~Vol.~XX, No.~XX, XX~2025}
	{Shell \MakeLowercase{\textit{et al.}}: A Sample Article Using IEEEtran.cls for IEEE Journals}
	\maketitle
	
	\vspace{-2mm}
	\begin{abstract}
		This paper investigates a design framework for sparse fluid antenna systems (FAS) enabling high-performance direction-of-arrival (DOA) estimation, particularly in challenging millimeter-wave (mmWave) environments. By ingeniously harnessing the mobility of fluid antenna (FA) elements, the proposed architectures achieve an extended range of spatial degrees of freedom (DoF) compared to conventional fixed-position antenna (FPA) arrays. This innovation not only facilitates the seamless application of super-resolution DOA estimators but also enables robust DOA estimation, accurately localizing more sources than the number of physical antenna elements. We introduce two bespoke FA array structures and mobility strategies tailored to scenarios with aligned and misaligned received signals, respectively, demonstrating a hardware-driven approach to overcoming complexities typically addressed by intricate algorithms. A key contribution is a light-of-sight (LoS)-centric, closed-form DOA estimator, which first employs an eigenvalue-ratio test for precise LoS path number detection, followed by a polynomial root-finding procedure. This method distinctly showcases the unique advantages of FAS by simplifying the estimation process while enhancing accuracy. Numerical results compellingly verify that the proposed FA array designs and estimation techniques yield an extended DoF range, deliver superior DOA accuracy, and maintain robustness across diverse signal conditions.
	\end{abstract}
	
	\begin{IEEEkeywords}
		Sparse fluid antennas (FA) array, direction-of-arrival (DOA) estimation, aligned received signals, misaligned received signals, polynomial root finding.
	\end{IEEEkeywords}
	
	\IEEEpeerreviewmaketitle
	\vspace{-2mm}
	\section{Introduction}
	\IEEEPARstart{T}{he} sixth generation (6G) wireless communication systems are anticipated to meet unprecedented requirements for high data rates, ultra-low latency, and spectral efficiency to support emerging applications such as immersive extended reality, holographic communications, and autonomous systems~\cite{New24}. Conventional multiple-input multiple-output (MIMO) architectures, despite their proven effectiveness in current networks, exhibit inherent limitations in scalability and adaptability for 6G scenarios. Fluid antenna systems (FAS) have recently emerged as a paradigm-shifting technology that addresses these fundamental limitations through physically reconfigurable antenna structures \cite{TWu20243, KKWong21, KKWong22}. By enabling dynamic port selection and position optimization within a constrained physical space, FAS fundamentally decouples the radiation characteristics from fixed hardware constraints, thereby facilitating adaptive spatial multiplexing and diversity exploitation \cite{Zhu-Wong-2024}. This architectural innovation enables FAS to achieve enhanced degrees of freedom, improved interference management, and superior channel utilization across diverse propagation environments. Recent investigations have demonstrated FAS's potential advantages in various aspects of wireless communications, including capacity enhancement, reliability improvement, and resource utilization optimization~\cite{XLai23, LaiX242, YaoJ251, YaoJ241, JYao2024, HXu23, Ghadi-2023, New-twc2023, NWaqar23, CWang24}.
	
	The concept of the fluid antenna (FA), sometimes referred to as a movable antenna (MA) \cite{Zhu-Wong-2024}, centers on the ability to dynamically reconfigure the position or activation of antenna elements to optimize communication performance. While some early prototypes explored fluidic implementations using conductive liquids within precision-controlled structures \cite{TWu20243}, the broader FA framework encompasses a variety of technologies—including electronically switchable arrays, reconfigurable metasurfaces \cite{BLiu25}, and spatially distributed antenna pixels \cite{Zhang25}—that achieve the same functional flexibility without any physical movement. This reconfigurability marks a key distinction from conventional fixed-position antenna (FPA) arrays, where element locations are permanently static. \textbf{By enabling dynamic control over which spatial points act as radiating elements, FA systems effectively expand the spatial degrees of freedom (DoF), creating a virtual aperture that surpasses the physical limitations of traditional antenna arrays} \cite{Ghadi-2023}. This spatial DoF expansion directly translates to enhanced communication performance metrics, including improved channel capacity, increased signal-to-noise ratio, and more effective interference mitigation \cite{YaoJ241}. While conceptually comparable to sparse antenna array implementations \cite{ref17, ref18, ref19, ref20}, FAS offer a critical operational advantage: they adaptively reconfigure spatial distributions in response to dynamic channel conditions, whereas sparse FPA arrays rely exclusively on predetermined element spacings that remain fixed throughout operation. This dynamic adaptability enables FA-based systems to achieve superior channel utilization efficiency without the mutual coupling limitations that typically constrain closely-spaced antenna elements, thereby presenting significant performance advantages across diverse wireless communication applications \cite{ref22,ref23,ref24,ref25,ref26,ref27,ref28,ref29,ref30}. 
	
	Beyond general communication enhancements, the expanded spatial DoF and virtual aperture extension afforded by FAS present particularly compelling advantages for direction-of-arrival (DOA) estimation. DOA estimation is a foundational task in array signal processing, supporting critical functions such as target localization in radar and sonar detection systems, channel estimation, and downlink precoding in wireless communication systems \cite{ref1,ref2,ref3}. Conventional DOA estimation techniques predominantly rely on FPAs. While numerous DOA estimation methods have been developed for FPA arrays, including linear prediction-based methods \cite{ref4,ref5}, subspace-based methods \cite{ref6,ref7}, sparse signal reconstruction-based methods \cite{ref8,ref9}, deep learning-based methods \cite{ref10,ref11}, and statistical inference-based methods \cite{ref12,ref13}, these approaches are inherently constrained by the fixed spatial DoF and limited flexibility of FPAs. These limitations often restrict their ability to achieve high-resolution DOA estimation, especially in complex scenarios with multiple closely spaced sources or underdetermined conditions. 
	
	In stark contrast, FAS, by strategically controlling element positions, can achieve substantially larger virtual apertures than physically possible with equivalent-sized FPA arrays, directly improving angular resolution capabilities and \textbf{achieving super-resolution}. Moreover, this controllable expansion of spatial diversity \textbf{enables stable operation in underdetermined DOA estimation scenarios}, allowing the accurate localization of more signal sources than the number of physical antenna elements -- a capability traditionally unattainable with conventional arrays of comparable physical dimensions. This ability to \textbf{achieve high-performance DOA estimation through innovative hardware design, which would otherwise necessitate algorithms of high computational complexity}, distinctly \textbf{reflects the unique characteristics and advantages of FAS}. Additionally, the flexible element spacing possible with FAS can be optimized to minimize mutual coupling effects that typically degrade estimation accuracy in densely packed arrays, while simultaneously maintaining the extended aperture necessary for high-resolution parameter estimation. These inherent advantages position FA-based DOA estimation as a promising solution to overcome fundamental limitations in conventional array processing techniques and unlock new potentials for high-performance spatial signal processing.

	However, realizing the full potential of FAS for DOA estimation is not without its hurdles. The practical implementation of FA-based DOA estimation must contend with several intrinsic challenges. Firstly, the very mobility that grants FAs their advantage is often constrained by physical limitations on the number of movements possible within the typically short observation times required for DOA estimation, where the direction information is assumed to be quasi-static \cite{ref34,ref35,ref36,ref37}. Secondly, translating these limited movements into a substantial and, crucially,  consecutive  expansion of spatial DoF, essential for many super-resolution algorithms, demands meticulous design of both the FA array geometry and its mobility strategy. Thirdly, particularly in mmWave bands targeted for future wireless systems, the presence of non-line-of-sight (NLoS) propagation paths can introduce significant multipath interference, complicating the accurate extraction of LoS signal parameters. Fourthly, the dynamic nature of the FA array necessitates robust and computationally efficient DOA estimation algorithms capable of processing the potentially time-varying array manifold. Finally, while FAS   offer the potential to mitigate mutual coupling through element spacing, this remains a design consideration to ensure the integrity of the received signals across different antenna positions.
	
	The aforementioned challenges are further compounded in scenarios involving misaligned received signals. To clarify this distinction, we formally define \textit{aligned received signals} as scenarios where the signals received at different FA positions are phase and time synchronized after each movement, as if they were received simultaneously by a larger, virtual array. This alignment enables direct combination of received samples from different positions to form a coherent virtual array with extended spatial response. Conversely, \textit{misaligned received signals} refer to scenarios where such phase and time synchronization cannot be maintained across FA movements, preventing direct signal combination due to independent signal variations at each position.
	
	While pilot-based communication systems may employ mechanisms to ensure signal alignment at the receiver after each FA movement, such alignment cannot be guaranteed in many critical DOA application contexts, including radar, sonar detection, or opportunistic signal-based collaborative positioning. In these non-cooperative or passive sensing scenarios, the signals received after successive FA movements are inherently misaligned due to the unknown and independent nature of the signal sources. This fundamental misalignment can severely degrade, or even nullify, the benefits accrued from spatial DoF expansion if not appropriately addressed, rendering accurate DOA parameter estimation significantly more complex and demanding more sophisticated array designs and signal processing techniques.
	
	To address these multifaceted challenges, this paper introduces novel FA array structures and corresponding mobility mechanisms   designed for DOA estimation under both aligned and misaligned received signal conditions. Complementing these hardware-centric innovations, we propose a low-complexity polynomial root-finding algorithm for robust DOA estimation, further enhanced by an effective method for detecting the number of line-of-sight (LoS) paths. Our simulation results demonstrate that these integrated solutions yield a significantly expanded range of consecutive spatial DoF, thereby enabling high-performance underdetermined DOA estimation with notable improvements in accuracy. The primary contributions of this work are detailed as follows:
	\begin{itemize} 
		\item \textbf{\textit{Optimized Sparse Array for Aligned Signals:}} Taking the scenario of aligned received signals and the limited number of movements into consideration, a sparse FA array structure consisting of two subarrays with large element spacing is first designed, and further optimized by applying different step size to the two subarrays, a significantly increased consecutive spatial DoF is achieved, which efficiently avoids the negative effects of array mutual coupling   and allows most   existing super-resolution estimators to be applied directly, finally facilitating the achievement of high-performance underdetermined DOA estimation. 
		\item \textbf{\textit{Generalized Hybrid Array for Misaligned Signals:}} A generalized FA array composed of an FPA array and an FA array with large element spacing is designed for misaligned received signals. It is demonstrated that by introducing a suitable PFA array, the array output data can be aligned in the second-order statistical domain and still provide increased consecutive spatial DoF. Meanwhile, this design is also applicable to scenarios with aligned received signals, offering higher universal applicability.    
		\item \textbf{\textit{LoS-Centric Closed-Form DOA Estimator:}} Considering the propagation characteristic of mm-wave signals, we couple the impact of  NLoS   signals into the  interference-plus-noise term, and then propose an approach for estimating the number of LoS paths/targets based on eigenvalue detection. Finally, a closed-form solution for DOA estimation is achieved by applying the polynomial root finding algorithm. Compared to the DOA estimation schemes that detect all paths, our scheme can provide significantly improved estimation accuracy and robustness. 
		\item \textbf{\textit{Comprehensive Performance Analysis and Validation:}} The performance in terms of estimation accuracy, computational complexity, capacity for underdetermined DOA estimation and the appropriate Cram\'{e}r-Rao Bound (CRB) for the considered scenarios are analyzed/provided. Extensive simulations under various FA array configurations are performed, which not only validate the effectiveness and superiority of the proposed solutions, but also facilitate their practical engineering implementation.
	\end{itemize}
	
	This paper is organized as follows. In Section II, the mm-wave signal model for FA array enabled DOA estimation is introduced, where both   aligned and misaligned received signals are considered. In Section III, two sparse FA array structures and their corresponding mobility mechanisms are designed and analyzed. In Section IV, we first introduce the LoS number detection based DOA estimation algorithms in detail, and then analyze their corresponding performance. Simulation results and main observations are provided in Section V, and conclusions are drawn in Section VI.
	
	{\emph{Notations:} Bold symbols in small letter and capital letter represent vectors and matrices, respectively. $a$ and $\mathcal{L}$ represent a scaler and a set, respectively. The superscripts ${\left(  \cdot  \right)^T}$, ${\left(  \cdot  \right)^H}$ and ${\left(  \cdot  \right)^*}$ stand for the transpose, conjugate transpose and conjugate, respectively. $\mathbb{E}\{  \cdot \} $ denotes the statistical expectation operation, ${\mathop{\rm diag}\nolimits} \left(  \cdot  \right)$ and ${\mathop{\rm blkdiag}\nolimits} \left(  \cdot  \right)$ are the diagonalization and block diagonalization operations, respectively. ${\mathop{\rm vec}\nolimits}(\cdot)$ represents the vectorization operator, which converts a matrix into a vector by stacking all of its columns on top of each other. $\otimes $ denotes the Kronecker product, and $\mathrm{spec}(\cdot)$ the eigenvalues of a matrix. $\mathrm{P_{eak}}(\cdot)$ returns the index corresponding to the first peak in the bracket. ${\bf{J}}_M$ denotes an $M\times M$ exchange matrix, and ${\bf{I}}_M$ an $M\times M$ identity matrix. For two sets $\mathcal{L}$ and $\mathcal{C}$, $\mathcal{L}\cup \mathcal{C}$ denote their union, and finally, $\mathcal{CN}({\bf 0},\sigma_n^2{\bf I}_M)$ indicates a complex Gaussian distribution with zero mean and variance $\sigma_n^2$.
		\vspace{-2mm}
		\section{Signal Model}
		\begin{figure}
			\centering
			\includegraphics[width=3.0in]{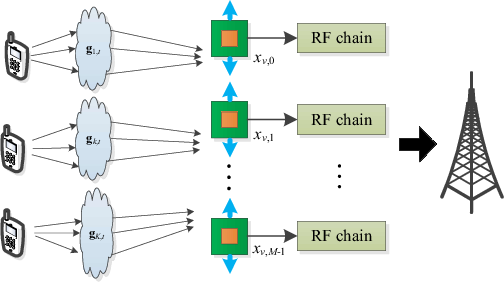}\\
			\caption{Illustration of FA array enabled DOA estimation system.}\label{Fig.antennas}
		\end{figure}
		Consider a mmWave narrowband system equipped with an $M$-element linear FA array \footnote{The term FA array hereafter denotes any array that contains movable radiating elements. When the received signals are misaligned, a subset of fixed-position antennas (FPAs) can be included to form the required consecutive spatial DoF; see Section~III for details.}, as illustrated in Fig.~\ref{Fig.antennas}.  The system simultaneously receives signals from $K$ single-antenna signal sources (referred to as targets throughout this paper), which can represent transmitting users in communication systems, radar targets, or any radiating objects whose DOAs are to be estimated.  Let $x_{v,m}$ denote the position of the $m$-th FA element after the $v$-th displacement, where $m\in\{0,\ldots ,M-1\}$ and $v\in\{0,\ldots ,G\}$.  FAS is assumed capable of performing $G$ such displacements within one channel coherence block.

		At mmWave frequencies, the propagation channel is characterized by its DoA and complex gain\cite{ref38,ref39,ref40}. Measurement studies have confirmed that angular parameters remain relatively stable, while small-scale fading coefficients change more rapidly\cite{ref41,ref42,ref43}. Throughout the $G$ consecutive movements, we therefore treat the path gains as quasi-static and neglect FA positioning errors, which typically remain below one-hundredth of the wavelength\cite{ref44}. Under these assumptions, the channel from the $k$-th signal source observed at the $t$-th snapshot after the $v$-th displacement can be expressed as
		\begin{equation}\label{1}
			\mathbf{h}_{k,v,t}=\sum_{l=1}^{L} g_{k,l,t}\mathbf{a}_v \left ( \theta _{k,l} \right ) , k\in[1,K],v\in[0,G],
		\end{equation}
		where $g_{k,1,t}$ and $g_{k,l,t}$ ($l\neq 1$) denote the complex Gaussian channel gains for the LoS and NLoS paths of the $k$-th target at the $t$-th snapshot, respectively. Here, $L$ represents the maximum number of paths considered for any target (or a common modeling parameter for the number of paths per target),  and $\mathbf{a} \left ( \theta _{k,l} \right )$ is the array manifold vector corresponding to the DOA $\theta _{k,l}$ of the $l$-th path of the $k$-th target, which can be expressed as 
		\begin{equation}\label{2}
			\mathbf{a}_v \left ( \theta _{k,l} \right ) =\left [e^{-jx_{v,0}\varpi_{k,l}},\dots , e^{-jx_{v,{M-1}}\varpi_{k,l}} \right ]^T,
		\end{equation}
		where $\varpi_{k,l}=2\pi \sin \theta _{k,l}/\lambda $, and $\lambda$ is the carrier wavelength.
		
		To facilitate subsequent analysis and system modeling, the uplink channel $\mathbf{h}_{k,v,t}$ can be expressed in a compact matrix-vector form as 
		\begin{equation}\label{3}
			\mathbf{h}_{k,v,t}=\mathbf{A}_{k,v}\mathbf{g}_{k,t},
		\end{equation}
		where
		\begin{equation}\label{4}
			\mathbf{A}_{k,v}=\left [\mathbf{a}_v \left ( \theta _{k,1} \right ),\dots, \mathbf{a}_v \left ( \theta _{k,L} \right )\right],
		\end{equation}
		\begin{equation}\label{5}
			\mathbf{g}_{k,t}=\left [g_{k,1,t},\dots, g_{k,L,t}\right ].
		\end{equation}
		Let $s_{k,v,t}$ denote the signal transmitted by the $k$-th target ($k\in[1,K]$) after the $v$-th movement at the $t$-th snapshot. We assume these signals are of equal power and are uncorrelated with each other. Under these assumptions, the received signal vector at the FA array, denoted by $\mathbf{y}_{v,t}$, can be expressed as
		\begin{eqnarray}\label{6}
			\begin{array}{l}
				\mathbf{y}_{v,t}  = \left [ \mathbf{h}_{1,v,t},\dots, \mathbf{h}_{K,v,t}  \right ] \mathbf{s}_{v,t}+\mathbf{n}_{v,t}\\
				\qquad = \left [\mathbf{A}_{1,v},\dots , \mathbf{A}_{K,v} \right ]\mathbf{G} \mathbf{s}_{v,t}+\mathbf{n}_{v,t} \\
				\qquad = \left [\mathbf{A}_{1,v},\dots , \mathbf{A}_{K,v}  \right ]\mathbf{\bar s}_{v,t} +\mathbf{n}_{v,t}= \mathbf{A}_v\mathbf{\bar s}_{v,t}+\mathbf{n}_{v,t},
			\end{array}
		\end{eqnarray}
		where $\mathbf{\bar s}_{v,t} =\mathbf{G} \mathbf{s}_{v,t}$ represents the equivalent received signal, $\mathbf{A}_v = \left[\mathbf{A}_{1,v},\dots , \mathbf{A}_{K,v}  \right ]$, $\mathbf{G}=\mathrm{blkdiag} \left (\left [ \mathbf{g}_{1,t},\dots , \mathbf{g}_{K,t} \right ]   \right ) $, and $\mathbf{n}_{v,t}\sim \mathcal{CN} \left (\mathbf{0}, \sigma_n^2\mathbf{I}_M  \right )$.
		
		At mmWave frequencies, the NLoS components of the channel gain are typically $5$ dB to $10$ dB weaker than the LoS  components \cite{ref45,ref46}. Leveraging this characteristic, we can distinguish between LoS and NLoS signal contributions. Consequently, the array output can be reformulated to emphasize the LoS components as follows
		\begin{eqnarray}\label{7}
			\begin{array}{l}
				{{\bf{y}}_{v,t}} = \sum\limits_{k = 1}^K {{{\bf{a}}_v}} \left( {{\theta _{k,1}}} \right){g_{k,1,t}}{s_{k,v,t}} + {{\bf{w}}_{v,t}}\\
				\qquad = {{\bf{A}}_{v,L}}{{{\bf{\bar s}}}_{v,L,t}} + {{\bf{w}}_{v,t}},
			\end{array}
		\end{eqnarray}
		where $\mathbf{A}_{v,L}=\left [\mathbf{a}_v\left (\theta_{1,1}\right ),\dots,\mathbf{a}_v\left (\theta_{K,1}\right )\right ]$ represents the array steering vector corresponding to the $K$ LoS paths, $\mathbf{\bar s}_{v,L,t}=\left [g_{1,1,t}s_{1,v,t},\dots ,g_{K,1,t}s_{K,v,t} \right ]^T$, and $\mathbf{w}_{v,t}$ denotes the interference-plus-noise term. Here, the interference component refers specifically to the NLoS signal contributions, and the complete expression is given by
		\begin{equation}\label{8}
			\mathbf{w}_{v,t}=\underbrace{\sum_{k=1}^{K} \sum_{l=2}^{L} \mathbf{a}\left (\theta_{k,l}\right )g_{k,l,t}s_{k,v,t}}_{\text{NLoS interference}}+\underbrace{\mathbf{n}_{v,t}}_{\text{additive noise}}.
		\end{equation}
		Due to the significant path loss experienced by NLoS paths and the Gaussian nature of the path gains, the interference-plus-noise term $\mathbf{w}_{v,t}$ can also be approximated as a complex Gaussian distribution \footnote{This assumption is commonly adopted in the literature, for instance, in reconfigurable intelligent surface (RIS) assisted localization \cite{ref47,ref48}, and is generally considered to align with practical scenarios. Our subsequent simulations will demonstrate that this approximation facilitates improved and more robust DOA estimation.}.  
		Consequently, the aggregated output of the FA array after $G$ movements, denoted by $\mathbf{y}_t$, is  expressed as 
		\begin{equation}\label{9}
			\mathbf{y}_t = \left[ {\begin{array}{*{20}{c}}
					\mathbf{y}_{0,t}\\
					
					\vdots \\
					\mathbf{y}_{G,t}
			\end{array}} \right] = \left[ {\begin{array}{*{20}{c}}
					\mathbf{A}_{0,L}&{}&{}\\
					{}& \ddots &{}\\
					{}&{}&\mathbf{A}_{G,L}
			\end{array}} \right]\left[ {\begin{array}{*{20}{c}}
					\mathbf{\bar s}_{0,L,t}\\
					\vdots \\
					\mathbf{\bar s}_{G,L,t}
			\end{array}} \right] + \mathbf{w}_t,
		\end{equation}
		where $\mathbf{w}_t=\left [ \mathbf{w}_{0,t},\dots , \mathbf{w}_{G,t}\right ]^T$.
		\begin{figure*}[t]
			\centering
			\includegraphics[width=6.6in]{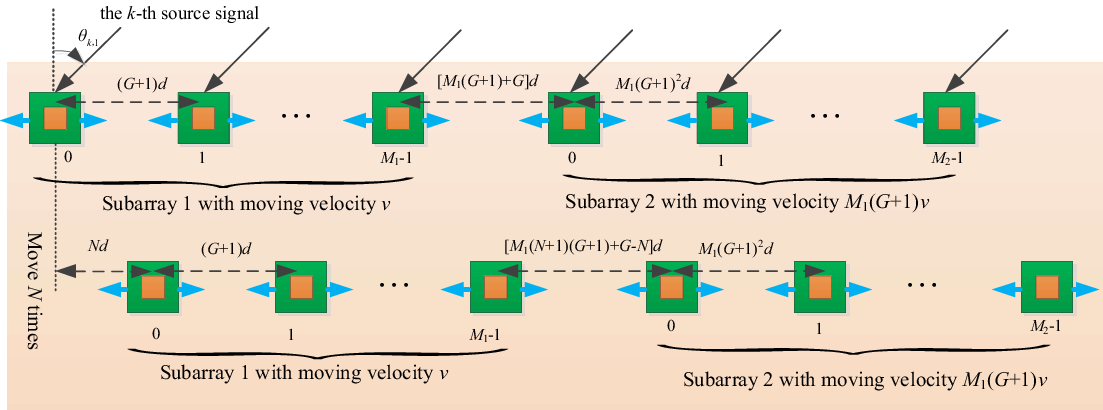} \vspace{-0mm}\\
			\caption{Sparse FA Array structure for aligned received signal scenarios.}\label{Fig.ARS} \vspace{-2mm}
		\end{figure*}
		By collecting $T$ snapshots, the FA array output can be expressed in matrix form as $\mathbf{Y}$, given by
		\begin{equation}\label{10}
			\mathbf{Y} =  \left[ {\begin{array}{*{20}{c}}
					\mathbf{A}_{0,L}&{}&{}\\
					{}& \ddots &{}\\
					{}&{}&\mathbf{A}_{G,L}
			\end{array}} \right]\left[ {\begin{array}{*{20}{c}}
					\mathbf{\bar S}_{0,L}\\
					\vdots \\
					\mathbf{\bar S}_{G,L}
			\end{array}} \right] + \mathbf{W},
		\end{equation}
		where $\mathbf{Y} =\left [ \mathbf{y}_1,\dots, \mathbf{y}_T \right ] $, $\mathbf{\bar S}_{v,L}=\left [\mathbf{\bar s}_{v,L,1},\dots , \mathbf{\bar s}_{v,L,T} \right ] $, and $\mathbf{W} =\left [ \mathbf{w}_1,\dots, \mathbf{w}_T \right ] $.
		
		In the particular case where the received signals are aligned at the receiver, i.e., $\mathbf{\bar S}_{L}=\mathbf{\bar S}_{0,L}=\dots =\mathbf{\bar S}_{G,L}$, \eqref{10} can be simplified as follows: 
		\begin{equation}\label{11}
			\mathbf{Y} = \mathbf{A}_L \mathbf{\bar S}_L+\mathbf{W},
		\end{equation}
		where $\mathbf{A}_L=\left [\mathbf{A}_{0,L}^T, \dots , \mathbf{A}_{G,L}^T\right ]^T=\left [\mathbf{\bar a}\left (\theta_{1,1}\right ),\dots,\mathbf{\bar a}\left (\theta_{K,1}\right )\right ]$.

		Building upon the signal model developed above, the remainder of this paper is organized as follows. In Section III, we will first introduce the design of sparse FA array structures and their corresponding mobility mechanisms. These designs are guided by the array manifold expression $\mathbf{A}_v$ and the concept of consecutive spatial  DoF, which will be formally defined. Subsequently, in Section IV, we will propose two efficient DOA estimation methods tailored to the scenarios described by \eqref{10} and\eqref{11}, respectively. 
		\section{Sparse FA Array Design}
		This section addresses the design of sparse FA array configurations and associated mobility mechanisms. The primary objective is to maximize the achievable consecutive spatial  DoF. We will develop designs tailored for scenarios involving both aligned and misaligned received signals. These array structures will serve as the foundation for the high-performance, low-complexity DOA estimation algorithms presented in the subsequent section. 
		
		Specifically, for aligned   signals, we design a sparse FA array structure consisting of two subarrays with large element spacing, where different step size is applied to the two subarrays to achieve significantly increased consecutive spatial DoF while avoiding array mutual coupling. For misaligned   signals, we propose a generalized FA array which is composed of an FPA array and an FA array with large element spacing, enabling the array output data to be aligned in the second-order statistical domain while still providing increased consecutive spatial DoF. The detailed design principles and performance analysis of these two array structures will be presented in the following subsections.
		\subsection{FA Array Designed for Aligned Received Signals}
		The FA array designed for aligned received signals consists of two uniform linear subarrays, denoted as subarray 1 and subarray 2, with $M_1$ and $M_2$ antenna elements respectively, such that the total number of antenna element is $M=M_1+M_2$. The design intricacies, illustrated in Fig.~\ref{Fig.ARS}, are pivotal for maximizing consecutive DoF. The inter-element spacing within the first subarray is set to $(G+1)d$, while the second subarray features a larger spacing of $M_1(G+1)^2d$. Here,  $d$ is the basic unit of movement, typically half the carrier wavelength. The initial separation distance between the two subarrays is $[M_1(G+1)+G]d$.
		
		The proposed design takes full advantage of the FA's mobility to create a dynamic sparse array configuration. {{By implementing different step sizes for the two subarrays, we can effectively expand the virtual array aperture while maintaining signal coherence.}} This unique approach not only enhances the spatial resolution but also provides { {better isolation between array elements, thereby reducing mutual coupling effects.}} The resulting configuration enables high-precision DOA estimation with improved robustness and efficiency.
		
		A key feature of this configuration is the differential mobility enabled by two independent controllers. Subarray 1 is moved by its controller at a nominal speed $v$, traversing a distance of $d$ per movement step. Concurrently, subarray 2 is moved by its dedicated controller at a significantly higher speed, $M_1(G+1)v$, covering a distance of $M_1(G+1)d$ in each movement step. For notational convenience, we define $\Delta_1 = G+1$ and $\Delta_2 = M_1(G+1)^2$. Subsequently, the coordinates of the elements in subarray 1 ($\mathcal{P}_N^{(1)}$) and subarray 2 ($\mathcal{P}_N^{(2)}$), as well as the overall array configuration ($\mathcal{P}_N$), after $N$ movements, are given by
		\begin{equation}\label{12}
			\mathcal{P}_N^{(1)}=\left \{ N, N+\Delta_1,\dots , N+(M_1-1)\Delta_1\right \} d,
		\end{equation}
		\begin{equation}\label{13}
			\mathcal{P}_N^{(2)} = \{ {\Delta _{13}^N},{\Delta _{13}^N} + {\Delta _2}, \ldots ,{\Delta _{13}^N} + ({M_2} - 1){\Delta _2}\}d,
		\end{equation}
		\begin{equation}\label{14}
			\mathcal{P}_N=\mathcal{P}_N^{(1)}\cup \mathcal{P}_N^{(2)},
		\end{equation}
		respectively, where ${\Delta _{13}^N} = (2{M_1} - 1){\Delta _1} + G + M_1 N{\Delta _1}$.
		
		\emph{Definition 1 (Consecutive Degrees of Freedom):}For a pair of FAs with coordinates $\ell_i d$ and $\ell_j d$, $\ell_i,\ell_j\in \mathbb{Z}$, the number of consecutive integers that $\ell_i-\ell_j$ can obtain is defined as the number of consecutive DoF.

		\emph{Proposition 1:} Consider the FA array designed for aligned received signals as described in Section III-A, comprising $M=M_1+M_2$ physical antennas and capable of $G$ movements ($\Delta_1 = G+1$). This configuration can achieve $2(M_1\Delta_1 - 1 + M_1M_2\Delta_1^2)+1$ consecutive spatial DoF. These DoF correspond to the continuous range of normalized spatial lags $[-(M_1\Delta_1 - 1 + M_1M_2\Delta_1^2), (M_1\Delta_1 - 1 + M_1M_2\Delta_1^2)]$. To maximize this number of consecutive DoF for a given total $M$, the optimal distribution of antennas between the two subarrays is:
		\begin{itemize}
			\item If $M$ is even: $M_1 = M_2 = M/2$;
			\item If $M$ is odd: $M_1 = (M+1)/2$ and $M_2 = (M-1)/2$ (or vice versa).
		\end{itemize}
		
		\begin{figure*}[t]
			\centering
			\includegraphics[width=6.6in]{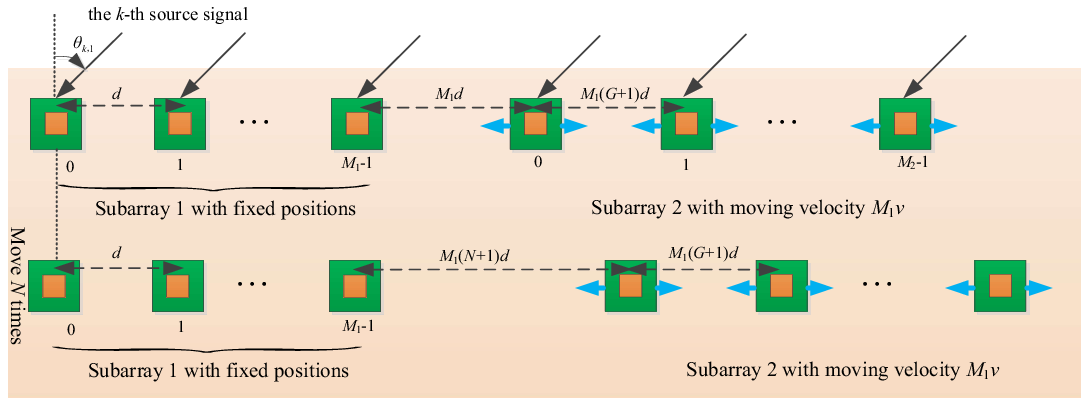} \vspace{-0mm}\\
			\caption{Sparse FA Array structure for misaligned received signal scenarios.}\label{Fig.MARS} \vspace{-2mm}
		\end{figure*}

		\begin{proof}
			Let $\mathcal{P}^{(1)}$ and $\mathcal{P}^{(2)}$ denote the complete sets of coordinates for all antennas in subarray 1 and subarray 2, respectively, after all possible movements. Specifically, $\mathcal{P}^{(1)}$ is the union of coordinates from all $G+1$ positions of subarray 1, i.e., $\mathcal{P}^{(1)}=\mathcal{P}_0^{(1)}\cup \mathcal{P}_1^{(1)}\cup\cdots \cup \mathcal{P}_G^{(1)}$, where $\mathcal{P}_N^{(1)}$ represents the coordinates of subarray 1 after the $N$-th movement. Similarly, $\mathcal{P}^{(2)}$ is defined as $\mathcal{P}^{(2)}=\mathcal{P}_0^{(2)}\cup \mathcal{P}_1^{(2)}\cup\cdots \cup \mathcal{P}_G^{(2)}$. After $G$ movements, the coordinate sets can be expressed as:
			
			\begin{equation}\label{15}
				\mathcal{P}^{(1)}=\left \{ \ell d | \ell=0,1,\dots ,M_1\Delta_1-1\right \},
			\end{equation}
			
			\begin{equation}\label{16}
				\begin{array}{l}
					\mathcal{P}^{(2)} = \{ \ell d|\ell  = {M_1}{\Delta _1} - 1 + {M_1}{\Delta _1},{M_1}{\Delta _1} - 1 + 2{M_1}{\Delta _1}\\
					\qquad\qquad \ldots ,{M_1}{\Delta _1} - 1 + {M_1}{M_2}\Delta _1^2\}.
				\end{array}
			\end{equation}
			
			In particular, $\mathcal{P}^{(1)}$ will generate $2M_1\Delta_1-1$ consecutive DoF within the range
			\begin{equation}\label{17}
				\mathcal{L}^{(1)}_a=[-M_1\Delta_1+1,M_1\Delta_1-1]
			\end{equation}
			
			$\mathcal{P}^{(2)}$ will generate $2M_2\Delta_1-1$ discontinuous DoF at an interval of $M_1\Delta_1$ within the range
			\begin{equation}\label{18}
				\mathcal{L}^{(2)}_a=[- (M_2\Delta _1 - 1)M_1 \Delta _1,(M_2\Delta _1 - 1)M_1 \Delta _1],
			\end{equation}
			and the differences between $\mathcal{P}^{(1)}$ and $\mathcal{P}^{(2)}$ will generate two $M_1M_2\Delta_1^2$ consecutive DoF within the ranges
			\begin{equation}\label{19}
				\mathcal{L}_c^{(1)} = [{M_1}{\Delta _1},{M_1}{\Delta _1} - 1 + {M_1}{M_2}\Delta _1^2] ,
			\end{equation}
			\begin{equation}\label{20}
				\mathcal{L}_c^{(2)} = [ - {M_1}{\Delta _1} + 1 - {M_1}{M_2}\Delta _1^2, - {M_1}{\Delta _1}],
			\end{equation}
			which directly yields $2({M_1}{\Delta _1} - 1 + {M_1}{M_2}\Delta _1^2)+1$ unique and consecutive spatial DoF, given by
			\begin{multline}\label{21}
				\mathcal{L}=\mathcal{L}^{(1)}_a\cup \mathcal{L}^{(2)}_a\cup \mathcal{L}^{(1)}_c\cup \mathcal{L}^{(2)}_c \\
				=[- {M_1}{\Delta _1} + 1 - {M_1}{M_2}\Delta _1^2,{M_1}{\Delta _1} - 1 + {M_1}{M_2}\Delta _1^2].
			\end{multline}
			
			Further define $f = 2({M_1}{\Delta _1} - 1 + {M_1}{M_2}\Delta _1^2)+1$, then, by replacing $M_2$ with $M-M_1$ and setting the derivative of $f$ with respect to $M_1$ equal to zero, we have
			\begin{equation}\label{22}
				M_1=\frac{M}{2} +\frac{1}{2\Delta_1}.
			\end{equation}
			
			Since $M_1$ is an integer, and $0<\frac{1}{2\Delta_1}\leq \frac{1}{4}$, provided that $G\geq 1$, the optimal values of $M_1$ and $M_2$ are
			\begin{equation}\label{23}
				\left\{ \begin{array}{l}
					{M_1} = {M_2} = \frac{M}{2},\;M\mathrm{\;is\;even}\\
					{M_1} = \frac{{M + 1}}{2},{M_2} = \frac{{M - 1}}{2},\;M\mathrm{\;is\;odd}
				\end{array} \right.
			\end{equation}
			and the maximum number of consecutive DoF are
			\begin{equation}\label{24}
				{f_{\max }} = \left\{ {\begin{array}{*{20}{l}}
						{M{\Delta _1} + \frac{{{M^2}\Delta _1^2}}{2} - 1,\;M\mathrm{\;is\;even}}\\
						{(M + 1){\Delta _1} + \frac{{({M^2} - 1)\Delta _1^2}}{2} - 1,\;M\mathrm{\;is\;odd}}.
				\end{array}} \right.
			\end{equation}
			
			This completes the proof of Proposition 1.
		\end{proof}
		\subsection{FA Array Designed for Misaligned Received Signals}
		For scenarios involving misaligned received signals, we propose a different FA array architecture termed as a generalized FA array. This configuration comprises two uniform linear subarrays with $M_1$ and $M_2$ antenna elements respectively ($M = M_1 + M_2$). The key distinction lies in the composition of the subarrays:
		\begin{itemize}
			\item Subarray 1 is an FPA array with a conventional inter-element spacing of $d$.
			\item Subarray 2 is an FA array whose elements are spaced $M_1(G+1)d$ apart.
		\end{itemize}
		
		Unlike conventional approaches that rely solely on algorithmic complexity to achieve high-resolution DOA estimation, the  { {proposed design leverages the intrinsic mobility of FA elements to create additional spatial degrees of freedom through physical movement.}} By carefully designing the array structure and movement mechanism, the system can achieve a virtual aperture and DoF expansion that would otherwise require much more complex signal processing algorithms.  { {This hardware-driven approach is a distinctive feature of  FAS, enabling high-performance DOA estimation with reduced computational burden and enhanced flexibility, especially in challenging misaligned signal scenarios.}}
		Due to the static nature of subarray 1, only subarray 2 is mobile, maneuvered by a single controller at a speed corresponding to $M_1v$. This design, illustrated in Fig.~\ref{Fig.MARS}, aims to generate substantial consecutive DoF even without signal alignment across movements. The effectiveness of this design is formalized in \textbf{\emph{Proposition 2}}.
		
		\textbf{\emph{Proposition 2:}} For the designed generalized FA array consisting of $M = M_1 + M_2$ antennas and allowing $G$ movements under the scenario of misaligned received signals, the achievable number of consecutive spatial DoF is $2(M_1 - 1 + M_1 M_2 \Delta_1) + 1$, corresponding to the range $[-M_1 + 1 - M_1 M_2 \Delta_1,\, M_1 - 1 + M_1 M_2 \Delta_1]$. To maximize the number of consecutive DoF for a given total $M$, the optimal allocation of antennas between the two subarrays is:
		\begin{itemize}
			\item When $M$ is even: $M_1 = M_2 = \frac{M}{2}$;
			\item When $M$ is odd: $M_1 = \frac{M+1}{2}$, $M_2 = \frac{M-1}{2}$ (or vice versa).
		\end{itemize}
		
		\begin{proof}
			After $G$ movements, we have
			\begin{equation}\label{25}
				\mathcal{P}^{(1)}=\left \{ \ell d | \ell=0,1,\dots ,M_1-1\right \}
			\end{equation}
			\begin{equation}\label{26}
				\begin{array}{l}
					\mathcal{P}^{(2)} = \{ \ell {M_1}d|\ell  = M_1 - 1 + M_1,M_1 - 1 + 2{M_1},\\
					\qquad\qquad \ldots ,M_1 - 1 +M_2\Delta_1 M_1\}.
				\end{array}
			\end{equation}
			
			It can be observed that $\mathcal{P}^{(1)}$ will generate $2M_1-1$ consecutive DoF within the range
			\begin{equation}\label{27}
				\mathcal{L}^{(1)}_a=[-M_1+1,M_1-1]
			\end{equation}
			while $\mathcal{P}^{(2)}$ will generate $2M_2\Delta_1-1$ discontinuous DoF at an interval of $M_1$ within the range
			\begin{equation}\label{28}
				\mathcal{L}^{(2)}_a=[- (M_2\Delta _1 - 1)M_1,(M_2\Delta _1 - 1)M_1]
			\end{equation}
			and the differences between $\mathcal{P}^{(1)}$ and $\mathcal{P}^{(2)}$ will generate two $M_1M_2$ consecutive DoF within the ranges
			\begin{equation}\label{29}
				\mathcal{L}_c^{(1)} = [M_1,M_2\Delta_1 M_1+M_1-1]
			\end{equation}
			\begin{equation}\label{30}
				\mathcal{L}_c^{(2)} =[-M_2\Delta_1 M_1-M_1+1,-M_1]
			\end{equation}
			which directly yields $2(M_1-1+M_1 M_2\Delta_1)+1$ consecutive DoF, given by
			\begin{multline}\label{31}
				\mathcal{L}=\mathcal{L}^{(1)}_a\cup \mathcal{L}^{(2)}_a\cup \mathcal{L}^{(1)}_c\cup \mathcal{L}^{(2)}_c \\
				=[-M_1+1-M_1 M_2\Delta_1,M_1-1+M_1 M_2\Delta_1].\;
			\end{multline}
			
			Define $f=2(M_1-1+M_1 M_2\Delta_1)+1$, and then by using the same partial differentiation way as in proposition 1, the optimal values of $M_1$ and $M_2$ can be obtained, which are the same as those in Eq. (23). Finally, the maximum number of consecutive DoF are
			\begin{equation}\label{32}
				{f_{\max }} = \left\{ {\begin{array}{*{20}{l}}
						{M-1 + \frac{{{M^2}\Delta _1}}{2},\;M\mathrm{\;is\;even}}\\
						{M + \frac{{({M^2} - 1)\Delta _1}}{2},\;M\mathrm{\;is\;odd}}.
				\end{array}} \right.
			\end{equation}
			This completes the proof of Proposition 2.
		\end{proof}
		
		\subsection{Insights of Two Different Arrays}
		Several remarks about the designed two FA arrays are given as follows to provide some interesting insights.
		
		\emph{Remark 1:} From \textbf{\textit{Proposition}} 1 and  \textbf{\textit{Proposition}} 2, we can observe that the number of consecutive spatial DoF increases significantly with both $G$ and $M$. This provides a crucial foundation for achieving high-performance underdetermined DOA estimation. It is worth noting that the minimum requirement for both array designs is two FA units (with $M_1=M_2=1$), which establishes the basic condition for implementing consecutive DoF extension.
		
		\emph{Remark 2:} The virtual array configuration resulting from the movement can be viewed as a specialized form of dynamic sparse nested array. However, our design differs fundamentally from conventional FP-based sparse nested arrays by explicitly incorporating the number of movements, $G$, into both the array structure and movement mechanism. This integration not only achieves higher spatial DoF but also provides enhanced flexibility. Particularly for aligned signal scenarios, the proposed design offers superior isolation between FA units in the two subarrays, effectively mitigating array mutual coupling effects. This represents a novel approach to improving the performance of wireless communication and signal processing systems.
		
		\emph{Remark 3:} While the first FA array (designed for aligned signals) achieves a significantly larger number of consecutive spatial DoF compared to the second array (designed for misaligned signals), the second array offers greater versatility by accommodating both aligned and misaligned signal scenarios. Therefore, when the characteristics of received signals are unknown, the second FA array design is recommended for its broader applicability.
		\section{DOA Estimation}
		To achieve robust DOA estimation while maintaining computational efficiency, we propose a closed-form DOA estimator based on LoS path number detection. This approach fully exploits the array output data while minimizing computational complexity. The detailed implementation and performance analysis of this estimator are presented in the following subsections.
		\subsection{LoS Path Number Detection Based DOA Estimation}
		For scenarios with aligned received signals, the array covariance matrix corresponding to the output $\mathbf{Y}$ is expressed as
		\begin{equation}\label{33}
			\mathbf{R}_Y=\mathbb{E}\left \{\mathbf{Y}\mathbf{Y}^H\right \} =\mathbf{A}_L\mathbf{R}_s\mathbf{A}_L^H+\mathbf{R}_W,
		\end{equation}
		where $\mathbf{A}_L$ represents the array manifold matrix for LoS paths, $\mathbf{R}_s$ denotes the signal covariance matrix, and $\mathbf{R}_W$ accounts for the interference-plus-noise term. The vectorized form of this covariance matrix is given by 
		\begin{equation}\label{34}
			\mathbf{z}=\mathrm{vec} \left (\mathbf{R}_Y\right ) =\mathbf{\tilde{A}}_L \mathbf{b} +\mathrm{vec} \left (\mathbf{R}_W\right),
		\end{equation}
		where  $\mathbf{\tilde{A}}_L=\left [\mathbf{\tilde{a}} \left ( \theta _{1,1} \right ),\mathbf{\tilde{a}} \left ( \theta _{2,1} \right ),\dots, \mathbf{\tilde{a}} \left ( \theta _{K,1} \right )\right]$ with its $k$-th column $\mathbf{\tilde{a}} \left ( \theta _{k,1} \right )=\mathbf{\bar a}^* \left ( \theta _{k,1} \right )\otimes \mathbf{\bar a} \left ( \theta _{k,1} \right )$. Here, $\mathbf{b}=[p_1,\ldots,p_K]^T$ is a column vector composed of the main diagonal elements of $\mathbf{R}_s$, which is defined as $\mathbf{R}_s=\mathbb{E}\{{\mathbf{\bar S}_L\mathbf{\bar S}_L^H}\}$.

		Define $\tau=(p - 1)MG + q$, $p,q\in[1,MG]$, then   the $\tau$-th element of $\mathbf{\tilde{a}}\left ( \theta _{k,1} \right )$ is  expressed as
		\begin{equation}\label{35}
			[\mathbf{\tilde{a}} \left ( \theta _{k,1} \right )]_{\tau}=\mathrm{exp} \left \{j(\tilde{x}_{p}-\tilde{x}_{q})\varpi _{k,1}  \right \},
		\end{equation}
		where $\tilde{x}_{p}$ and $\tilde{x}_{q}$ represent the virtual coordinates wrt the $p$-th element in $\mathbf{\bar a}^* \left ( \theta _{k,1} \right )$ and the $q$-th element in $\mathbf{\bar a}\left ( \theta _{k,1} \right)$, respectively.
		
		Based on Proposition 1, we have
		\begin{equation}\label{36}
			\left \{\tilde{x}_{p}-\tilde{x}_{q},p,q\in[1,MG]\right \} \subseteq \mathcal{L}.
		\end{equation}
		This relationship enables us to construct the following observation vector by rearranging the elements in $\mathbf{z}$
		\begin{equation}\label{37}
			\begin{array}{l}
				{\mathbf{r}} = {\mathbf{Cb}} + \mathbf{\bar w} = \left[ {{\bf{c}}\left( {{\theta _{1,1}}} \right), \ldots ,{\bf{c}}\left( {{\theta _{K,1}}} \right)} \right]{\bf{b}} +\mathbf{\bar w}\\
				= \left[ {\begin{array}{*{20}{c}}
						{{e^{ - j\Delta d{\varpi _{1,1}}}}}& \cdots &{{e^{ - j\Delta d{\varpi _{K,1}}}}}\\
						\vdots & \cdots & \vdots \\
						1& \cdots &1\\
						\vdots & \cdots & \vdots \\
						{{e^{j\Delta d{\varpi _{1,1}}}}}& \cdots &{{e^{j\Delta d{\varpi _{K,1}}}}}
				\end{array}} \right]\times  \left[ {\begin{array}{*{20}{c}}
						{{p_1}}\\
						{{p_2}}\\
						\vdots \\
						{{p_K}}
				\end{array}} \right] + \mathbf{\bar w}
			\end{array}
		\end{equation}
		where $\Delta={M_1}{\Delta _1} - 1 + {M_1}{M_2}\Delta _1^2$, and $\mathbf{\bar w}$ represents the transformed interference-plus-noise term derived from $\mathbf{W}$.
		For scenarios with misaligned received signals, we obtain a series of sub-covariance matrices, expressed as
		\begin{equation}\label{38}
			{{\bf{R}}_g} = {{\bf{A}}_{g,L}}{{\bf{R}}_{g,L}}{\bf{A}}_{g,L}^H + {{\bf{R}}_{w,g}},
		\end{equation}
		where ${{\bf{R}}_{g,L}}$ and ${{\bf{R}}_{w,g}}$ denote the sub-covariance matrices of received signals and noise corresponding to the array output data after the $g$-th movement.Under the assumption of equal power for transmitted signals and additive noise, we have  ${{\bf{R}}_{0,L}}={{\bf{R}}_{1,L}}=\cdots={{\bf{R}}_{G,L}}$, and ${{\bf{R}}_{w,0}}\approx{{\bf{R}}_{w,1}}\approx\cdots\approx{{\bf{R}}_{w,G}}$. By incorporating the results from \textbf{\textit{Proposition 2}}, we can construct an observation vector ${\mathbf{r}}$ with the same structure as \eqref{37}, but with a modified  $\Delta$ (i.e., $\Delta=M_1-1+M_1 M_2\Delta_1$) for the misaligned signal scenario.
		
		Next, we divide $\mathbf{r}$ into $\Delta+1$ overlap sub-vectors, and then construct a $(\Delta+1)\times(\Delta+1)$ matrix, written by
		\begin{multline}\label{39}
			\mathbf{R}_r=\left [\mathbf{J}_{\Delta+1}[\mathbf{r}]_{1:\Delta+1},\dots , \mathbf{J}_{\Delta+1}[\mathbf{r}]_{\Delta+1:2\Delta+1}  \right ] \\
			=\mathbf{\bar C}\mathbf{R}_s\mathbf{\bar C}^H+\mathbf{R}_{W'},\qquad\qquad\qquad\qquad\qquad\qquad
		\end{multline}
		where $\mathbf{R}_{W'}$ is constructed by the elements of $\mathbf{\bar w}$, and $\mathbf{\bar C}=\left[ \mathbf{\bar c}\left( \theta _{1,1} \right),\mathbf{\bar c}\left( \theta _{2,1} \right), \ldots ,\mathbf{\bar c}\left( \theta _{K,1} \right) \right]$, with its $k$-th column $\mathbf{\bar c}\left( \theta _{k,1} \right)$ given by
		\begin{equation}\label{40}
			\mathbf{\bar c}\left( \theta _{k,1} \right)=\left [0,e^{-jd\varpi_{k,1}},\dots , e^{-j\Delta d\varpi_{k,1}} \right ]^T.
		\end{equation}
		
		It can seen that $\mathbf{\bar C}\mathbf{R}_s\mathbf{\bar C}^H$ holds the same structure as that obtained with an $(\Delta+1)$-element PFA based uniform linear array (ULA) under the LoS propagation scenarios, allowing most of existing ULA based solutions can be applied directly. To make a good trade off between computational complexity and DOA estimation accuracy, the  eigenvalue detection based polynomial root finding algorithm is exploited here.
		
		Based on the uncorrelated characteristics of channel gains at different time blocks and the expression of $\mathbf{w}_{v,t}$ in Eq. (8), $\mathbf{R}_{W'}$ can be written as
		\begin{equation}\label{41}
			\mathbf{R}_{W'}=\mathbf{\tilde{C}}\mathbf{R}_{s'}\mathbf{\tilde{C}}^H+\sigma_n^2\mathbf{I}_{\Delta+1}
		\end{equation}
		where $\mathbf{\tilde{C}}$ and $\mathbf{R}_{s'}$ represent the transformed array manifold matrix and received signal power matrix wrt NLoS components, respectively.
		
		According to the properties of eigenvalues for sample array covariance matrices, the following relationship holds\cite{ref49},\cite{ref50}:
		\begin{equation}\label{42}
			\begin{array}{l}
				\mathrm{spec}  ( \mathbf{\hat R}_r) = (\alpha _1 \ldots ,\alpha _K,\beta _1, \ldots ,\beta _{(L - 1)K},\underbrace {0, \ldots ,0}_{1 \times (\Delta+1 - LK)})\\
				+ \sigma _n^2(\underbrace {1, \ldots ,1}_{1 \times MG}) = \sigma _n^2(\underbrace {{\alpha _1}/\sigma _n^2 + 1, \ldots ,{\alpha _K}/\sigma _n^2 + 1}_{{\rm{LoS\;eigenvalues}}})\\
				+ \sigma _n^2(\underbrace {{\beta _1}/\sigma _n^2 + 1, \ldots ,{\beta _{(L - 1)K}}/\sigma _n^2 + 1}_{{\rm{NLoS\; eigenvalues}}}) + \sigma _n^2(\underbrace {1,1, \ldots ,1,1}_{{\rm{noise\;eigenvalues}}})\\
				= (\rho _1,\rho _2, \ldots ,\rho _{\Delta+1})
			\end{array}
		\end{equation}
		where $\mathbf{\hat R}_r$ represents the sample covariance matrix estimated from $T$ snapshots, ${\alpha _i}$ and $\beta_j$ denote the true eigenvalues of $\mathbf{\bar C}\mathbf{R}_s\mathbf{\bar C}^H$ and $\mathbf{\tilde{C}}\mathbf{R}_{s'}\mathbf{\tilde{C}}^H$, respectively. Due to the significant power difference between LoS and NLoS path gains, a distinct gap exists between $\{\rho _1,\ldots,\rho _K\}$ and $\{\rho _{K+1},\ldots,\rho _{LK}\}$ regardless of the received  SNRs. This characteristic enables us to estimate the number of LoS paths through the following approach 
		\begin{equation}\label{43}
			\hat K = \mathrm{P_{eak}}\left( f_k={{\rho _k}/{\rho _{k + 1}}} \right),k = 1, \ldots ,\Delta.
		\end{equation}
		
		\emph{Remark 4:} Accurate estimation of the number of sources/paths is a crucial prerequisite for DOA estimation. While most existing methods rely on the Akaike's information criterion (AIC) or the minimum description length (MDL) criterion\cite{ref51},\cite{ref52},\cite{ref53}, these approaches are primarily designed for detecting the total number of paths. Although our proposed LoS path number estimation algorithm is relatively simple, it proves more effective in achieving accurate and robust DOA estimation performance, as demonstrated in subsequent simulations.
		
		With the estimated number of LoS paths $\hat K$, we can perform eigenvalue decomposition (EVD) on $\mathbf{\hat R}_r$ to obtain
		\begin{equation}\label{44}
			\mathbf{\hat R}_r=\mathbf{U}_{s'}\mathbf{\Sigma}_{s'}\mathbf{U}_{s'}^H +\mathbf{U}_{W'}\mathbf{\Sigma}_{W'}\mathbf{U}_{W'}^H
		\end{equation}
		where the column of $\mathbf{U}_{s'}$ and $\mathbf{U}_{W'}$ represent the  eigenvectors corresponding to
		the signal subspace and generalized interference-plus-noise subspace bases, respectively. $\mathbf{\Sigma}_{s'}$ and $\mathbf{\Sigma}_{W'}$ are diagonal matrices containing the largest $K$ eigenvalues and remaining small eigenvalues of $\mathbf{\hat R}_r$, respectively.
		
		Using the obtained $\mathbf{U}_{W'}$, we can construct the root polynomial as
		\begin{equation}\label{45}
			f\left (z  \right ) =\mathbf{p}^T(z^{-1})\mathbf{U}_{W'}\mathbf{U}_{W'}^H\mathbf{p}(z)
		\end{equation}
		where $\mathbf{p}(z)=\left [1,z,z^2,\dots ,z^\Delta  \right ]^T $. By identifying the $K$ solutions closest to the unit circle in \eqref{45}, the DOA angle can be estimated as
		\begin{equation}\label{46}
			\hat\theta_{k,1}=\arcsin \left ( \frac{\lambda }{2\pi d}\angle \hat z_k  \right ) ,k=1,2,\dots ,K.
		\end{equation}
		\subsection{Performance Analysis and Appropriate  Cram\'{e}r-Rao Bound}
		\subsubsection{Estimation Accuracy}
		The proposed solutions rely on the polynomial root finding algorithm to achieve DOA estimation. According to the derivation in \cite{ref54}, the mean squared error (MSE) of DOA estimation satisfies
		\begin{equation}\label{47}
			\mathbb{E}\left\{ {{{\left| {{\theta _{k,1}} - {{\hat \theta }_{k,1}}} \right|}^2}} \right\} = {\left( {\frac{\lambda }{{2\pi d\cos {\theta _{k,1}}}}} \right)^2}\frac{{\mathbb{E}\left\{ {{{\left| {{z_k} - {{\hat z}_k}} \right|}^2}} \right\}}}{{2\left( {\Delta  + 1} \right)}}
		\end{equation}
		which is proportional to the polynomial root estimation error and inversely proportional to $\Delta$ (or precisely $M$ and $G$). This implies that the DOA estimation performance improves as $M$ and $G$ increase. However, for the polynomial root estimation error, it is impacted by various factors, including the number of snapshots, NLoS components, and additive noise, making its impact on DOA estimation performance more complex. In particular, under the limited snapshot condition, the received signal power when calculating the sub-covariance matrix after each movement is difficult to remain consistent. In the process of constructing ${\mathbf{r}}$ through elements rearrangement, not only is a subset of elements used, but also a misalignment bias is introduced. Further combining the unavoidable NLoS path components, a fixed bias term will appear in the polynomial root estimation error, which is independent of noise variance. As a result, the MSE ``saturates" at this bias point even as the noise variance tends to zero or SNR is sufficiently high, as intuitively verified later by simulations.
		\subsubsection{Computational Complexity}
		The proposed solutions are closed-form ones, whose main computational complexity arises from the construction of the covariance matrix and eigenvalue decomposition (EVD). For  aligned   signals, the proposed method constructs one $MG\times MG$ covariance matrix with $T$ snapshots, and performs the EVD on a $(\Delta+1)\times(\Delta+1)$ matrix, therefore it roughly requires $\mathcal{O}(M^2G^2T+\frac{3}{4}(\Delta+1)^3)$. While for misaligned   signals, the proposed method constructs $G$ sub-covariance matrices of size $M\times M$ with $T$ snapshots, and still performs the EVD on a $(\Delta+1)\times(\Delta+1)$ matrix, which totally require $\mathcal{O}(GM^2T+\frac{3}{4}(\Delta+1)^3)$. In comparison with the spectrum search based methods, such as   MUSIC   \cite{ref6} and   RARE   \cite{ref7}, the proposed solutions avoid the spectral peak search and offer  better computational efficiency. In addition, it is necessary to point out that $\Delta$ in misaligned   signals   is much smaller than that for aligned   signals, therefore, the DOA estimator for misaligned   signals is computationally more efficient.
		
		\emph{Remark 5:} The performance analysis reveals a fundamental trade-off in FA array design. While increasing $M$ and $G$ theoretically enhances DOA estimation accuracy by expanding the spatial DoF, practical constraints such as limited snapshots and NLoS effects introduce a performance "ceiling" characterized by the MSE saturation phenomenon. This underscores the importance of our array structures that efficiently exploit the available spatial aperture with minimal FA elements and movements. For time-sensitive applications, the aligned signal design offers superior accuracy, whereas the misaligned signal design provides better computational efficiency without compromising robustness. System designers should weigh these considerations based on specific application requirements and environmental conditions.
		
		\subsubsection{Capacity for Underdetermined DOA Estimation}
		The theoretical foundation for our system's ability to handle underdetermined scenarios stems directly from subspace theory. Analysis of the EVD process reveals a fundamental constraint: at least one eigenvector from the constructed covariance matrix must be reserved for spanning the interference-plus-noise subspace $\mathbf{U}_{W'}$. This constraint establishes an upper bound on the number of detectable LoS paths or targets, specifically $K\leq \Delta$. 
		
		The significance of this bound becomes apparent when examining the maximum achievable value of $\Delta$, which can reach $\frac{f_{\max}-1}{2}$. For the aligned signal scenario in particular, we have demonstrated that $f_{\max} \geq 2M+1$ when $G \geq 1$ and $M \geq 2$. This relationship reveals a critical insight: our proposed FA array designs can identify more targets than the number of physical antennas employed, thereby achieving true underdetermined DOA estimation. This capability represents a substantial advantage over conventional fixed-position array systems, which are fundamentally limited by their physical antenna count.
		
		The theoretical capacity for underdetermined estimation derived above is comprehensively validated through our simulation results, which demonstrate reliable estimation of multiple targets using significantly fewer physical antenna elements. This confirms that the proposed FA array architecture efficiently exploits the spatial aperture expansion enabled by controlled antenna mobility. 
		
		\begin{figure*}[t]
			\centering  
			\vspace{-0.35cm} 
			\subfigtopskip=2pt 
			\subfigbottomskip=-2pt 
			\subfigcapskip=-5pt 
			\subfigure[]{\includegraphics[width=3.3in]{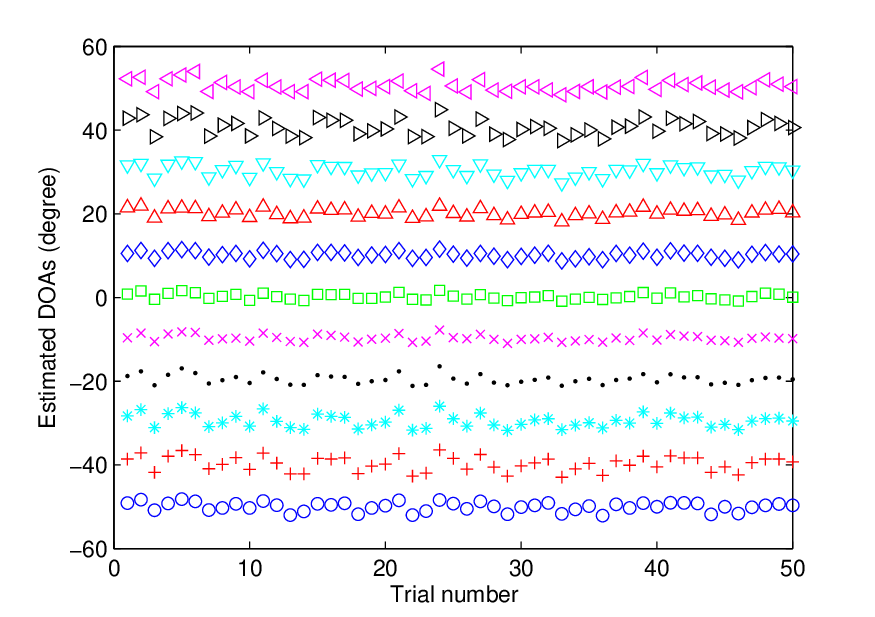}\label{fig3.a}}\hspace{-2mm}  
			\subfigure[]{\includegraphics[width=3.3in]{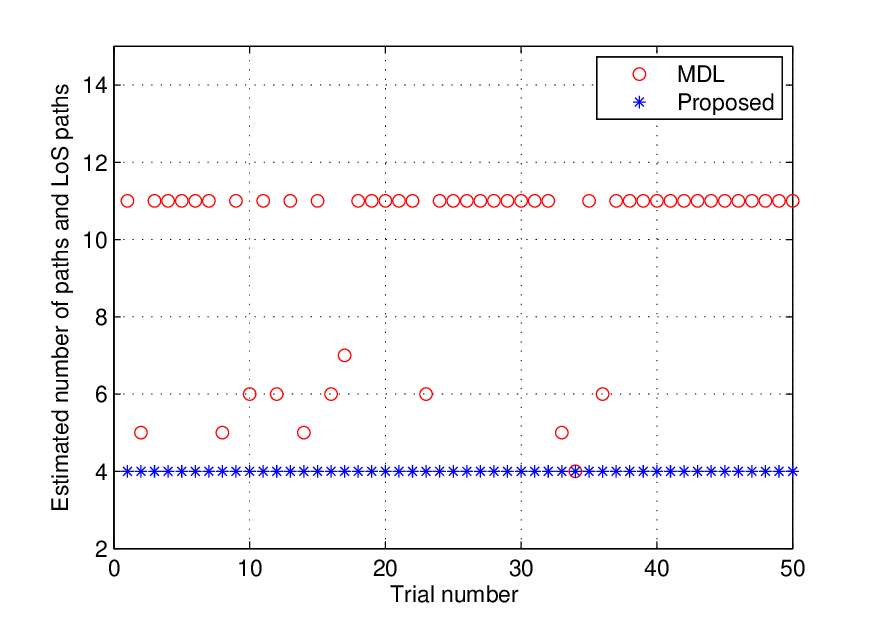}\label{fig3.b}}\hspace{-2mm}
			
			\subfigure[]{\includegraphics[width=3.3in]{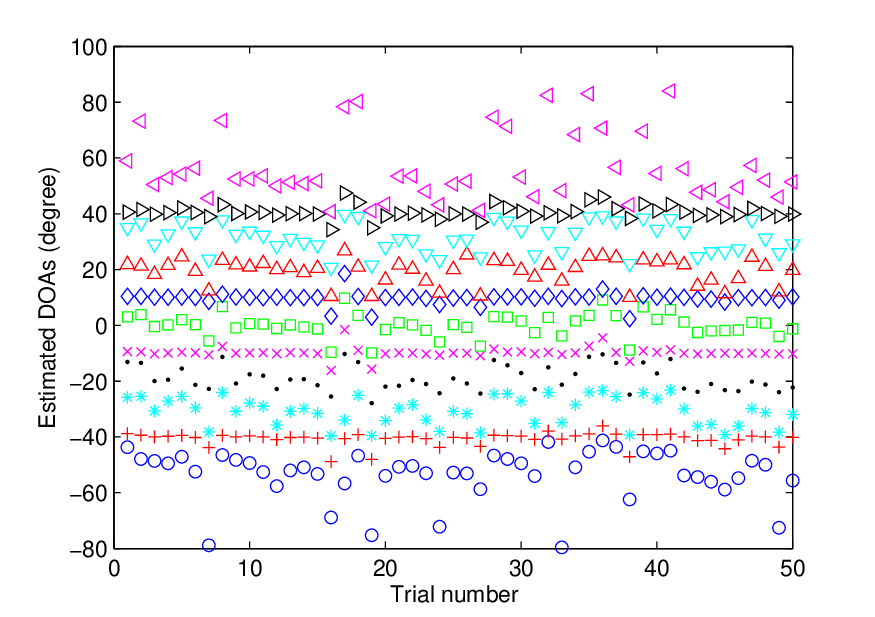}\label{fig3.c}}\hspace{-2mm}
			\subfigure[]{\includegraphics[width=3.3in]{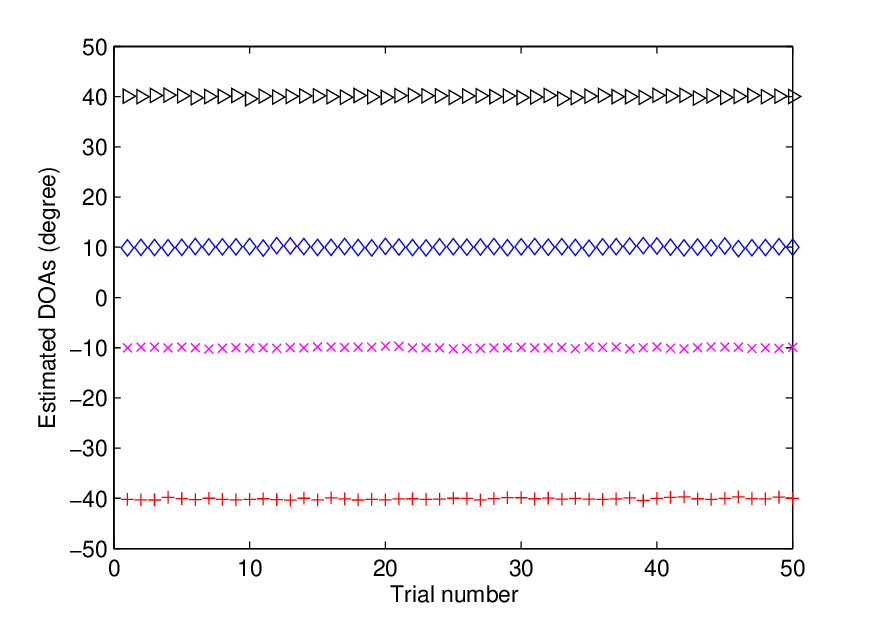}\label{fig3.d}}\hspace{-2mm}
			
			\caption{Scatter  plots of estimated DOA angles and LoS number for aligned received signals: (a) DOA estimation with eleven LoS targets; (b) LoS number estimation with four LoS paths and seven NLoS paths; (c) DOA estimation result employing the number of all paths; (d) DOA estimation result employing the number of LoS paths.}
			\label{level}
		\end{figure*}
		\subsubsection{Appropriate Cram\'{e}r-Rao Bound (CRB)}
		The Cram\'{e}r-Rao Bound (CRB) provides a fundamental theoretical limit on estimation accuracy, serving as a critical benchmark for evaluating the performance of our proposed methods. A distinguishing feature of our approach is that it focuses solely on DOA estimation for LoS paths, while treating NLoS components as part of the interference-plus-noise term $\mathbf{W}$. This modeling choice is justified by the severe attenuation experienced by NLoS components in mmWave propagation environments.
		
		To facilitate mathematical tractability while maintaining model fidelity, we approximate $\mathbf{W}$ as interference drawn from an isotropic complex Gaussian distribution -- an  approach that has been successfully employed in related works \cite{ref47,ref48}. For the theoretical performance analysis of underdetermined DOA estimation, we adopt Jansson's CRB framework \cite{ref55,ref56} with uncorrelated prior, which is specifically designed for scenarios where the number of targets exceeds the number of physical antenna elements. Within this framework, the $(a,b)$-th element of the Fisher Information Matrix (FIM) is expressed as:
		\begin{equation}\label{48}
			[\mathbf{F} ]_{a,b}=T\mathrm{vec}^H\left ( \frac{\partial \mathbf{R} _Y}{\partial \boldsymbol{\varsigma}_a } \right )\mathbf{\Psi} \mathrm{vec}\left ( \frac{\partial \mathbf{R} _Y}{\partial \boldsymbol{\varsigma}_b } \right ),
		\end{equation}
		where $\mathbf{\Psi} =\mathbf{R} _Y^{-T}\otimes \mathbf{R} _Y^{-T}$ represents the weighted matrix, and $ \boldsymbol{\varsigma}=\left [\theta_{1,1},\dots ,\theta_{K,1},\mathbf{p} \right ]$ denotes the parameter vector comprising DOA angles and power parameters.
		
		By defining the gradient matrices $\mathbf{D}_\theta =\left [\mathrm{vec}(\partial\mathbf{R}_Y/\partial\theta_{1,1}),\dots ,\mathrm{vec}(\partial\mathbf{R}_Y/\partial\theta_{K,1}) \right ]$ and $\mathbf{D}_p =\left [\mathrm{vec}(\partial\mathbf{R}_Y/\partial p_{1}),\dots ,\mathrm{vec}(\partial\mathbf{R}_Y/\partial p_{K}) \right ]$, the FIM can be compactly expressed as:
		\begin{equation}\label{49}
			\mathbf{F}=T\left [\mathbf{D}_\theta\;  \mathbf{D}_p\right ]^H \mathbf{\Psi} \left [\mathbf{D}_\theta\;  \mathbf{D}_p\right ].
		\end{equation}
		
		A crucial theoretical result from \cite{ref56} indicates that the FIM is invertible when $2K\leq 2\Delta+1$. This condition not only reinforces the underdetermined estimation capability of our design, but also establishes the mathematical validity of the CRB expression. Under these conditions, the closed-form CRB for DOA estimation is given by:
		\begin{equation}\label{50}
			\mathrm{CRB}_\theta =\frac{1}{T}  \left \{\mathbf{D}_\theta^H\mathbf{\Psi}^{1/2}\mathbf{P}^\perp _ {\mathbf{\Psi}^{1/2}\mathbf{D}_p} \mathbf{\Psi}^{1/2} \mathbf{D}_\theta  \right \}^{-1},
		\end{equation}
		where $\mathbf{P}^\perp _\mathbf{X} =\mathbf{I} -\mathbf{X} (\mathbf{XX} ^H)^{-1}\mathbf{X} ^H$ denotes the orthogonal projection operator.
		\begin{figure*}[h]
			\centering  
			\vspace{-0.35cm} 
			\subfigtopskip=2pt 
			\subfigbottomskip=-2pt 
			\subfigcapskip=-5pt 
			\subfigure[]{\includegraphics[width=3.2in]{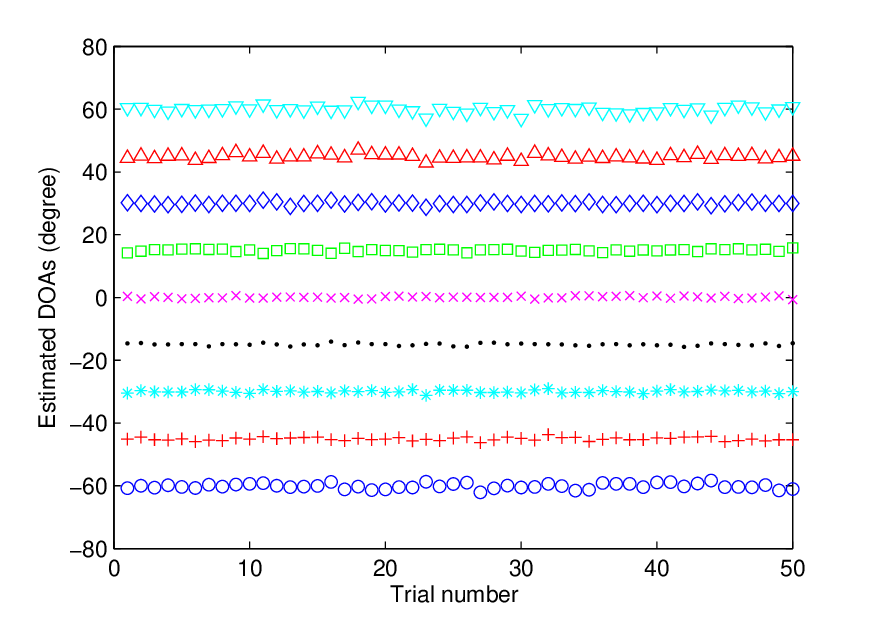}\label{fig4.a}}\hspace{-2mm}  
			\subfigure[]{\includegraphics[width=3.2in]{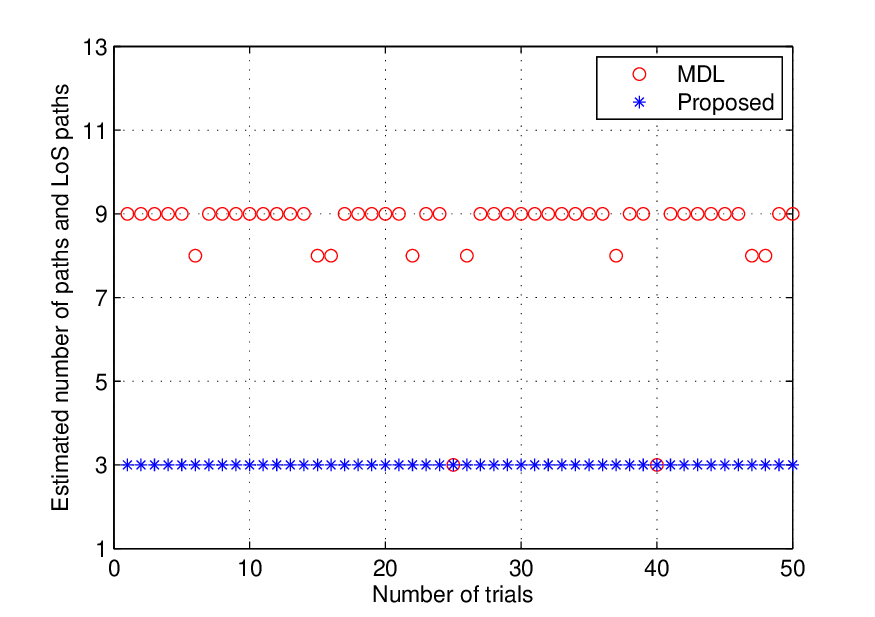}\label{fig4.b}}\hspace{-2mm}
			
			\subfigure[]{\includegraphics[width=3.2in]{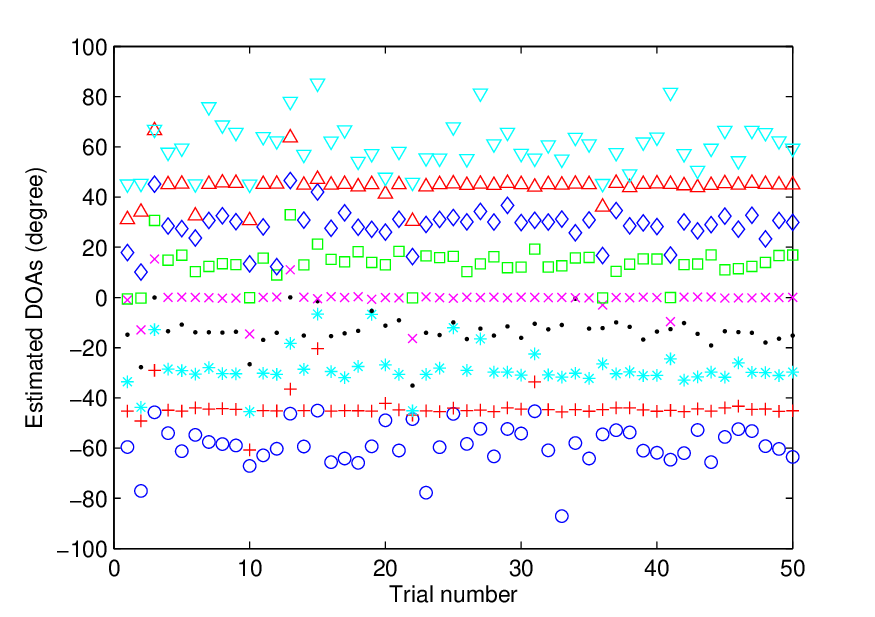}\label{fig4.c}}\hspace{-2mm}
			\subfigure[]{\includegraphics[width=3.2in]{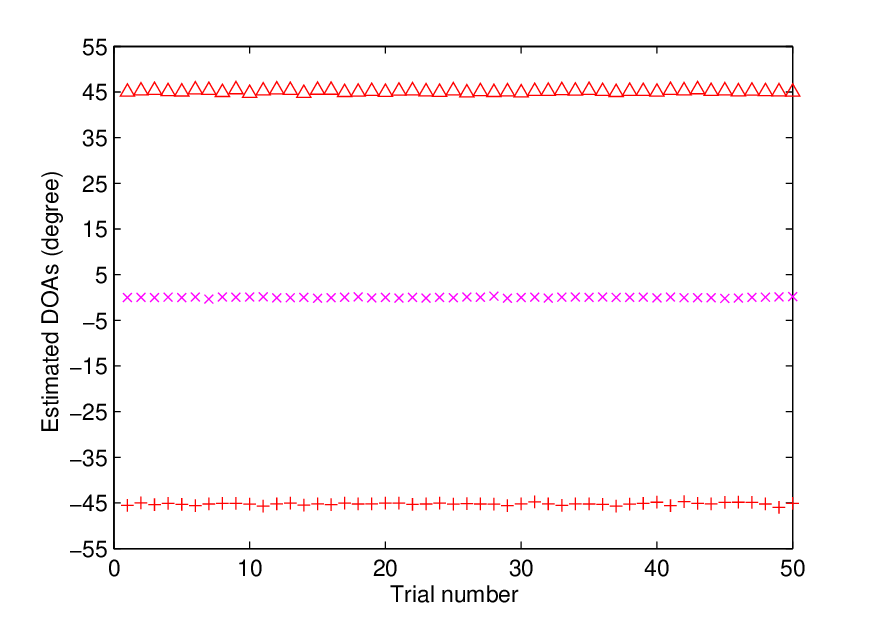}\label{fig4.d}}\hspace{-2mm}
			
			\caption{Estimated DOA angles and LoS number of 50 independent trials for misaligned received signals: (a) DOA estimation with nine LoS targets; (b) LoS number estimation with three LoS paths and six NLoS paths; (c) DOA estimation result employing the number of all paths; (d) DOA estimation result employing the number of LoS paths.}
			\label{level}
		\end{figure*}

		\begin{figure*}[t]
			\centering  
			\vspace{-0.35cm} 
			\subfigtopskip=2pt 
			\subfigbottomskip=-2pt 
			\subfigcapskip=-5pt 
			\subfigure[]{\includegraphics[width=3.2in]{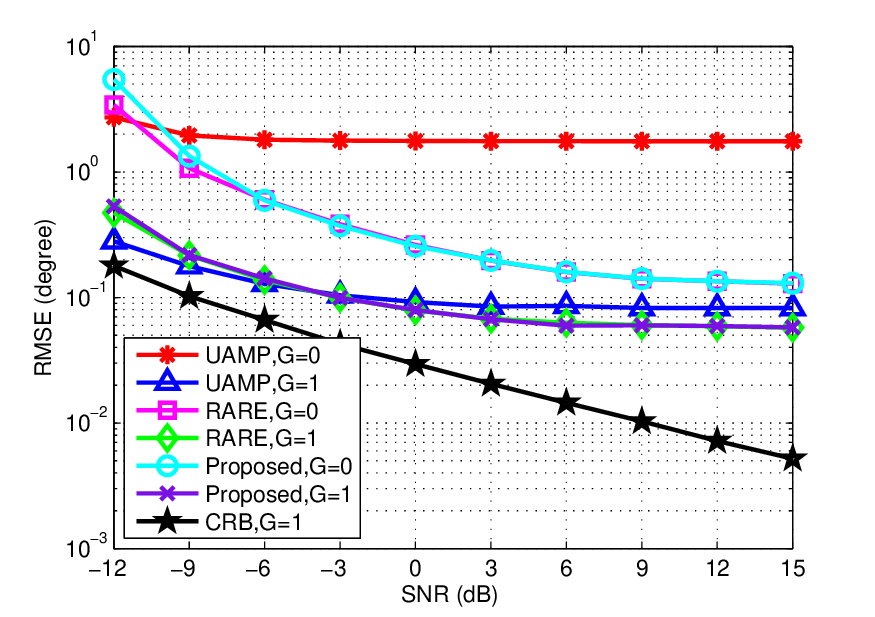}\label{fig5.a}}\hspace{-2mm}  
			\subfigure[]{\includegraphics[width=3.2in]{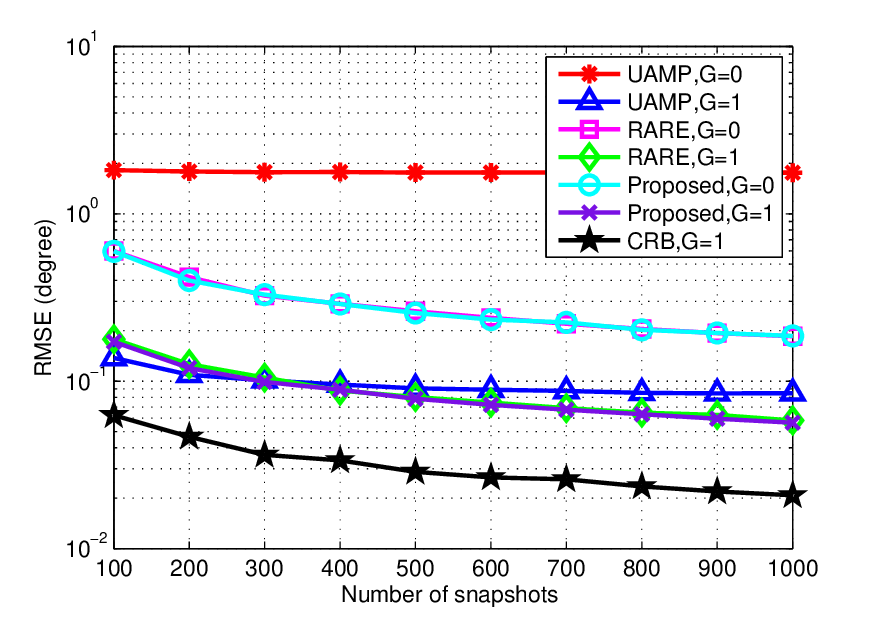}\label{fig5.b}}\hspace{-2mm}
			\caption{RMSE of DOA estimation versus SNR and the number of snapshots under the scenario of aligned received signals, $\theta_{1,1}=-20.3^\circ, \theta_{2,1}=10.7^\circ$, and DOAs wrt NLoS paths are within the range of $[\theta_{k,1}-5^\circ,\theta_{k,1}+5^\circ]$: (a) versus SNR; (b) versus the number of snapshots.}
			\label{level}
		\end{figure*}
		
		\begin{figure*}[t]
			\centering  
			\vspace{-0.35cm} 
			\subfigtopskip=2pt 
			\subfigbottomskip=-2pt 
			\subfigcapskip=-5pt 
			\subfigure[]{\includegraphics[width=3.2in]{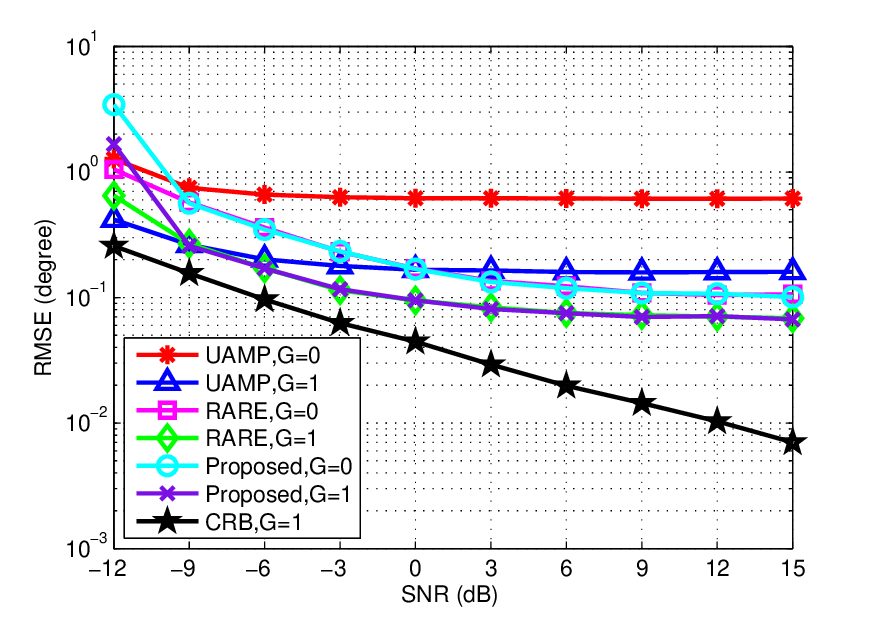}\label{fig5.a}}\hspace{-2mm}  
			\subfigure[]{\includegraphics[width=3.2in]{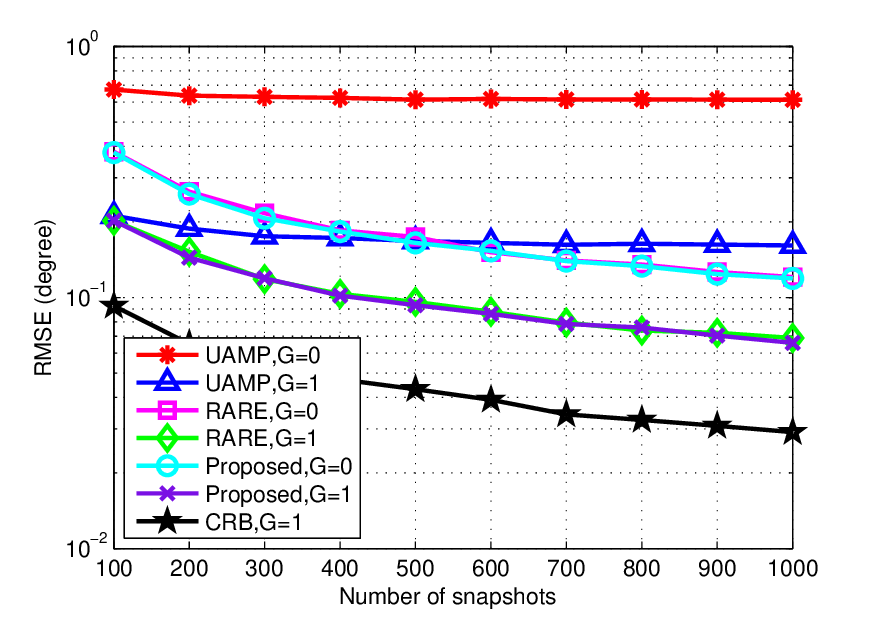}\label{fig5.b}}\hspace{-2mm}
			\caption{RMSE of DOA estimation versus SNR and the number of snapshots under the scenario of misaligned received signals, $\theta_{1,1}=-20.3^\circ, \theta_{2,1}=10.7^\circ$, and DOAs wrt NLoS paths are within the range of $[\theta_{k,1}-5^\circ,\theta_{k,1}+5^\circ]$: (a) versus SNR; (b) versus the number of snapshots.}
			\label{level}
		\end{figure*}
		\section{Simulation Results}
		In this section, the DOA estimation performance of the proposed two solutions is assessed, and compared with that of the subspace based RARE   \cite{ref7}, sparse signal reconstruction based unitary approximate message passing (UMAP)   \cite{ref9}, as well as the CRB. Meanwhile, the MDL criterion   \cite{ref52} is also selected to compare the LoS path number estimation performance of the proposed eigenvalue detection based estimator. For DOA estimation, the root mean square error (RMSE) obtained by averaging 500 independent Monte-Carlo trials is adopted to assess the performance of different methods.
		
		In the first simulation, we demonstrate the ability of the proposed method to achieve underdetermined DOA estimation, as well as the superiority of the developed LoS path number detection based DOA estimator in the scenario of aligned received signals. The simulation result with 50 trials is shown in Fig, 4, where the number of movements is $G=1$ with basic moving unit $d$ equaling to the half the carrier wavelength, the initial coordinates of three FAs are $\{0, 2, 7\}d$, SNR=10 dB, and the number of snapshots $T$ is set to 1000 . In Fig. 4(a), eleven LoS targets with DOAs uniformly distributed within the range of $[-50^\circ,50^\circ]$ are considered. It can be seen that the proposed solution can successfully estimate eleven targets with only three FAs, efficiently validating the capacity to handle underdetermined DOA estimation problems. In Figs. 4(b) to 4(d), four LoS paths with their DOAs $\{-40^\circ,-10^\circ,10^\circ,40^\circ\}$ and seven NLoS paths with their DOAs $\{-50^\circ,-30^\circ,-20^\circ,0^\circ,20^\circ,30^\circ\}$ are considered, the amplitude of channel gain wrt NLoS components is 10 dB weaker than that wrt LoS components. Fig. 4(b) show the LoS path number estimation result, from which we can see that the MDL estimates the total number of paths in most cases, and it is even not robust. In contrast, the proposed scheme can effectively estimate the number of LoS paths with very robust results. Figs. 4(c) and 4(d) further show the DOA estimation results rely on the total number of paths and the number of LoS paths, respectively. It can be observed that the LoS path-based estimators provide significantly improved and more robust DOA estimation results, whereas the total path-based estimator exhibits considerable bias. This highlights the effectiveness and superiority of the proposed solution.
		
		In the second simulation, we continue to evaluate the ability of the proposed solution to achieve underdetermined DOA estimation and estimation superiroty under the scenario of misaligned received signals. The simulation configurations are as follows: the total number of antennas is four, which comprises of two FPAs with their coordinates $\{0, 1\}d$ and two FAs with their initial coordinates $\{3, 7\}d$, SNR, $G$ and the number of snapshots $T$ are set to 10 dB, 1 and 500, respectively. In Fig. 5(a), nine LoS targets with DOAs uniformly distributed within the range of $[-60^\circ,60^\circ]$ are considered,  respectively. In Fig. 5(b) to 5(d), three LoS paths with their DOAs $\{-45^\circ,0^\circ,45^\circ\}$ and six NLoS paths with their DOAs $\{-60^\circ,-30^\circ,-15^\circ,15^\circ,30^\circ,60^\circ\}$ are considered. The other simulation conditions are the same with the first simulation. It can be seen that the proposed method can not only estimate nine LoS targets with only four antennas, but also provide improved and more robust estimation result, again demonstrating its great capacity for underdetermined DOA estimation and superiority for improved DOA estimation in mm-wave transmission environments.
		
		In the third simulation, we evaluate the DOA estimation performance under different SNRs and snapshots for the scenario of aligned received signals. Two LoS paths with their DOAs $\theta_{1,1}=-20.3^\circ, \theta_{2,1}=10.7^\circ$ are considered. For each LoS path, there are two corresponding NLoS paths with corresponding DOAs randomly distributed within the range of $[\theta_{k,1}-5^\circ,\theta_{k,1}+5^\circ]$ and amplitudes are attenuated by 10 dB. The initial coordinates of three FAs are $\{0, 2, 7\}d$, and $G$ is set to 0 and 1. In Fig. 6(a), the number of snapshots is fixed at 500, and SNR varies from -12 dB to 15 dB, whereas in Fig. 6(b), SNR is fixed at 0 dB, the number of snapshots varies from 100 to 1000 in steps of 100. As can be seen from the simulation results, the RMSE of DOA estimations decreases as the increase of the number of movements $G$, SNR and the number of snapshots, showing the advantage of the designed FA array structure and mobility mechanism efficiently. On the other hand, it can be observed that the proposed polynomial root finding based solution can provide a better estimation performance in comparison with the UMAP when $T>200$, and almost achieve the same performance with the RARE method. However, as analyzed before, the proposed method belongs to the close-form one, which is computationally more efficient than the RARE method. In addition, it can be further seen that the performance of the proposed solution suffers from ``saturation phenomenon", leading a clear gap between the proposed method and CRB in high SNRs, which is consistent with the performance analysis in Section IV.
		
		In the last simulation, we show the RMSEs versus SNR and the number of snapshots under the scenarios of misaligned received signals. The simulation conditions are the same with the third simulation, except that four antennas with two FPAs are located at $\{0, 1\}d$ and two FAs initially are located at $\{3, 7\}d$. Note that the UMAP is established on the time-domain output of array, but the array structure and mobility mechanism we designed do not introduce additional spatial DoF in misaligned scenarios. Therefore, for comparison, we clarify here that the results obtained by UMAP are based on the model with aligned received signals at the coordinates $\{0, 1, 2, 5\}d$ under the four FAs. As can be seen in Fig. 7, the proposed solution again performs better than the UMAP method, and achieves low complexity and almost the same estimation accuracy with the RARE method, demonstrating the effectiveness of superiority of the proposed solution.
		
		\vspace{-2mm}
		\section{Conclusion}
		In this paper, we proposed fluid antenna array designs and estimation techniques that offer substantial advantages for DOA estimation in mmWave systems. By intelligently exploiting antenna mobility and sparse array configurations, the framework significantly increases the spatial degrees of freedom, enabling high-resolution DOA estimation even in underdetermined scenarios. Unlike conventional fixed arrays, the proposed approach supports both aligned and misaligned signal conditions while maintaining low complexity and enhanced robustness. These capabilities position fluid antenna systems as a promising hardware-driven solution for next-generation wireless sensing and communication systems, where adaptability, precision, and efficient spatial processing are critical.
		%

	\end{document}